# Galaxy Mass Distributions from Rotation Speeds by Closed-Loop Convergence

Kenneth F. Nicholson (Caltech Alumni)
email: knchlsn@alumni.caltech.edu

**Abstract**

Given the dimensions (including thickness) and rotation speeds for an axisymmetric galaxy, a closed-loop iterative method is designed to find the total mass, density, and surface-mass density (SMD) distributions. Density accuracy of course depends on the accuracy of the thickness data, but SMD and total mass are as good as the inputs of rotation speeds vs. radius if reasonable estimates are made for thickness. The effects of an exponential "atmosphere" are included.

## 1. Introduction

Over the past 30 years or so the determination of SMD vs. radius has mostly depended on the use of dark-matter spherical shells of varying density, designed to allow an approximate match of the computed and measured rotational speeds. Because physical evidence of the existence of the dark-matter spheres has been elusive, P. J. E. Peebles (1993) was led to comment: "Discovery of the source of the dark matter, or explaining why the Newtonian mechanics used to infer its existence has been misapplied, has to be counted as one of the most exciting and immediate opportunities in cosmology today." Binney and Tremaine (1987, B & T used after this) say "There is no well-established example of a Keplerian region in any galaxy rotation curve, even those that extend to radii large enough to contain essentially all of the galaxy's light. Consequently, *there is no spiral galaxy with a well-determined total mass.*" (Italics are theirs.) They then draw what I believe is the wrong conclusion: They say " The simplest explanation of these results is that other spiral galaxies, like our own, possess massive dark halos that extend to larger radii than the optical disks, a conclusion first stated by Freeman (1970)." Figure 1 shows how this process works. A recent example with many references is that of Jimenez, Verde, and Oh (2002).

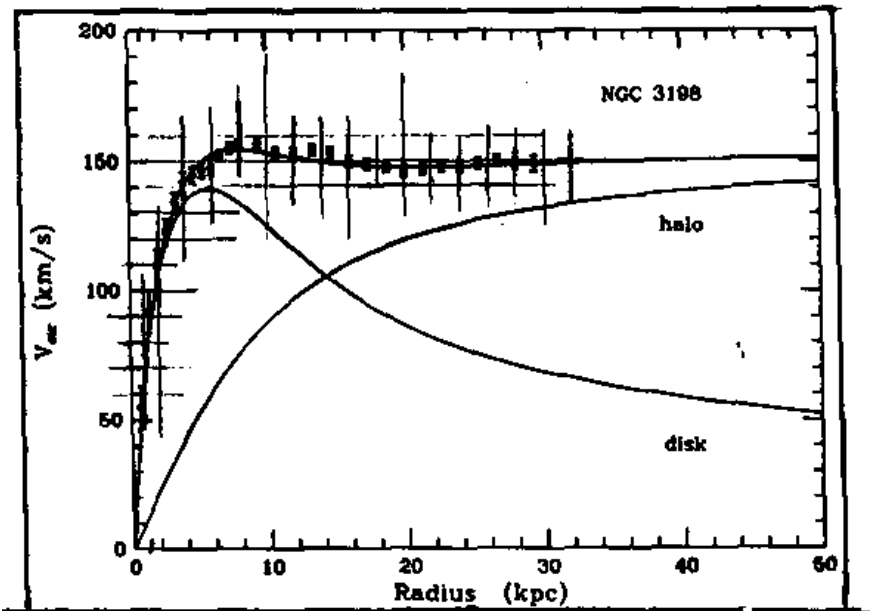

Figure 1. Use of a dark-matter halo with an exponential disk to approximate
the measured rotation curve. The squares of the speeds are added. By
Van Albada et al (1985)

I believe the correct conclusion should be that the mathematics are wrong. Milgrom (2002) having invented MOND, has tried to eliminate the dark-matter spheres by changing the constants in Newton's law with distance. But to distances out to the edge of our own planetary system, NASA and JPL have shown that Newton's law works very well. So here I use Newton's law with no changes, and apply it in the usual engineering way, assuming the galaxies have definite thickness distributions, and are made up of ordinary matter.

The forward problem, finding the rotation speeds from the mass distribution,, should be easy using a computer. It should be no more difficult than finding the mass, inertias, and center of gravity of a large vehicle. But of course it isn't easy. Physically, for a single isolated ring of dust-like material, a test mass passing radially through it from the inside should first show an increase in acceleration outward from the center, have the acceleration smoothly change signs as it passes through the ring, then proceed asymptotically to the Kepler result. The mathematics most in use cannot show this smooth effect, and cause trouble with divisions by zero. However, all divisions by zero can be avoided, smooth acceleration changes through an isolated ring can be demonstrated, and rotation curves are easily generated for complete galaxies. High confidence in the method, equations, and computer coding is achieved by comparisons with theory.

Using a computer it will be shown that the acceleration of a test mass, slightly outside the mid radius of a "dusty" isolated ring, is the same as if all the ring mass was concentrated at the center. Although some will say it's obvious, it is not easy to show mathematically.

B & T show two methods using potential theory to solve the forward problem. One method uses elliptic integrals, the other uses Bessel and Hankel functions, and both methods are applied to a very thin disk. In theory both methods can be used with an arbitrary distribution of SMD, but in practice no one does this because of the difficult and tedious math involved, and troubles with divisions by zero. A solution of the second method that has been widely used in the community (see Figure 1) assumes an exponential form of the SMD versus radius. The constants in this exponential form ($\Sigma_o$ and $R_d$) can be set by using correlations of SMD and galaxy surface brightness (mass to light ratio). The notation here uses subscripts to agree with the source, but in most of this paper computer notation is used to avoid subscripts, ie each term can be defined with several letters.

For SMD = $\Sigma = \Sigma_o \, e^{-2y}$

$$v^2(R) = 4 \pi G \, \Sigma_o \, R_d \, y^2 \, [\, I_o(y) \, K_o(y) - I_1(y) \, K_1(y)\,] \qquad (1)$$

where $y \equiv R / (2 R_d)$, and I and K are modified Bessel functions of the first and second kind

note that $R_d$ is not the maximum galaxy radius here, but merely a shaping constant

B & T also show a solution for the reverse problem, that is finding SMD vs radius knowing the rotational speeds vs radius (their equation 2-174). That solution is wrong on two counts. It uses $dv^2/dR$ over the entire radius of the galaxy, but from experiment v is often constant to large radii. It also uses the elliptic integral K in the form K(1) that is infinite. To my knowledge no one uses this equation.

I can only speculate on how equation (1) and the dark-matter spheres came to be linked. Perhaps because the galaxies seemed to show surface light intensity that falls off exponentially from the center, it might have been assumed that equation (1) was nearly correct for the galaxy disks. Therefore something else had to be causing the differences found in experiment. By adjusting the density of spherical shells centered on a galaxy, their gravitational effects could be added to those shown by equation (1) to get a much better comparison with experiment. Because there was no physical evidence of these spherical shells, the concept of dark-matter halos was born. It is said that the dark matter was implied by the gravitational effects.

However there is no physical or mathematical reason to do this. Galaxies can have an infinite number of types of mass distributions, few of them closely exponential. The correct application of Newton's law can find the correct rotational speeds for any mass distribution. No dark-matter spheres are implied. Also there is no need for stars or light to be present. For gravitational and dynamic effects, galaxies can be small, starless (because of low densities), and made of ordinary stuff, like sand. If astronomers want to find more of the "missing mass" needed to balance out the universe, perhaps they should consider lots of small starless galaxies.

The reverse problem, finding mass distribution from rotation speeds, is more difficult, and so far as I can tell has never been done correctly. As pointed out above, equation 2-174 in B & T cannot be right. If the galaxy is assumed to be made up of rings of constant SMD, the SMD of each ring must be adjusted so that the measured rotation speed at each radius can be matched. Yet the speed at any radius is affected by all the rings. Providing a way can be found to get a correction method that converges, this is a good place to use an iterative approach. With a computer, the process can continue until all speeds are matched to a specified accuracy. The concept is simple: start with any density distribution, compute the forward problem to get the resulting rotation speeds, compare these with measured data, correct the densities in a direction to reduce the errors, repeat until the errors are all less than some specified value. Doing this however requires more art than science.

It turns out that if the experimental data are faired and interpolated to have results at the outer edge of each ring, the error there (computed minus measured) can be used nicely to correct the speed ( by changing that ring density) without unduly changing the speeds elsewhere in the galaxy. The size of the corrections (ie setting the gain) must be adjusted to converge fairly fast yet allow a stable process. Judicious use of limits on the corrections are used to help the stability. Another concern is where to locate the rod representing the mass of each fundamental particle ($h \times rd\theta \times dr$, h=thickness). For the rod to be at the centroid of a segment with inner radius $r_i$, it should be at $r_i + 2/3\, dr$ at $r_i \to 0$, and move to $r_i + dr/2$ as $r \to \infty$. However when checking by generating the rotation profile from a given mass distribution, then using that profile to find the original mass distribution, it was found to work better to have it slightly more toward the mid radius of each ring.

Another problem is how to deal with the way density varies with z ( perpendicular to the galaxy plane). The computing becomes too complex if each segment is assumed to have a large-body part (more or less elliptic in z) and a gas/dust "atmosphere" part (more or less exponential in z). So the fundamental segment has to be simplified to have constant density in z yet be equivalent to this complex model.

The convergence process continues until computed and measured rotation speeds match within vmax E−6, where the measured data are an interpolated value for each radius. Because the final result of this process is the output of a forward problem, the same high confidence carries over to the reverse solution.

It was noticed while doing this problem that the equations and answers can be made dimensionless. The dimensionless forms allow the presentation of data for many galaxies on the same plots, and allows easy comparisons of galaxies, much like the test results for aircraft. The computing here is done in dimensioned form, and the data are presented in dimensionless form. Because SMD gets very large near the center, it is sometimes presented as $\log_{10}$ (SMD). However I have chosen to present it as the product of radius and SMD. This allows the data to be easily read across the full radius, and the slope at the center is proportional to the SMD there.

## 2. Method

The assumptions are that the galaxies are axisymmetric and have smooth thickness distributions, that the rotation speeds are measured at the center plane, and that the galaxy has a definite outer rim where the last speeds are measured. The speeds are more likely to be measured near the surface of the galaxy rather than the center plane. The effects of the speed measurement location will be small at large radii, but should have some effect near the center. Also any material outside the defined rim will cause the readings near the rim to be low. As seen in figure 2, the thickness of the "large-body" region is far from smooth, but the outer surface of the gas/dust region is fairly good. The gas/dust outer surface thickness is about 6 times that of the large-body thickness for this example. For most galaxies like that of figure 2, or even some with a central bulge, a reasonable thickness distribution and maximum radius can be defined.

If in doubt, a "Milky Way" thickness distribution, similar to that of figure 2, is used. "Zero" (ie very thin) thickness should not be used because that causes low values for both SMD and total mass. The effects of thickness have been previously shown by Nicholson (2000). For a given mass, thickness reduces the acceleration of a test mass at the rim of a galaxy, as compared with a thin disk.

The first step is to find a constant-density fundamental segment that has the same mass (or SMD) and is equivalent to a model segment of a real galaxy. The distribution of large-body and gas/dust densities used here was purely a guess based on looking at figure 2, and to simplify things it is assumed unchanged with radius. The model segment was designed to have twice as much gas/dust as large bodies, with the density of the "atmosphere" part constant to the surface of the large-body region, then declining exponentially to 1% of the initial value at z= +/−3 × (large-body thickness) Beyond that the atmosphere is neglected.

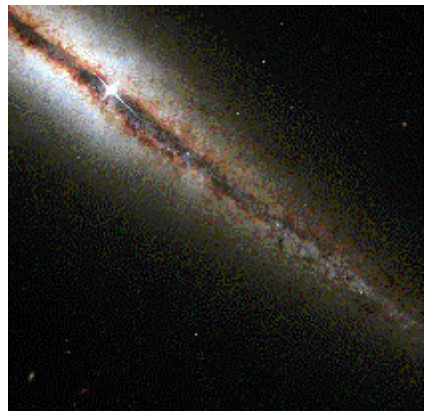

Figure 2.   NGC 4013 on edge,  Hubble

The equivalent and model segments are in figure 3. The thickness and corresponding density of the equivalent segment was found by comparing results for a single ring. In use, the model characteristics are measured or estimated, the equivalent thickness found for computing, equivalent densities vs r found from the measured speeds, then the model densities found from that. SMD does not change. In general the model-segment characteristics will change with radius, but they don't here, and the effect on SMD and even densities is considered second order.

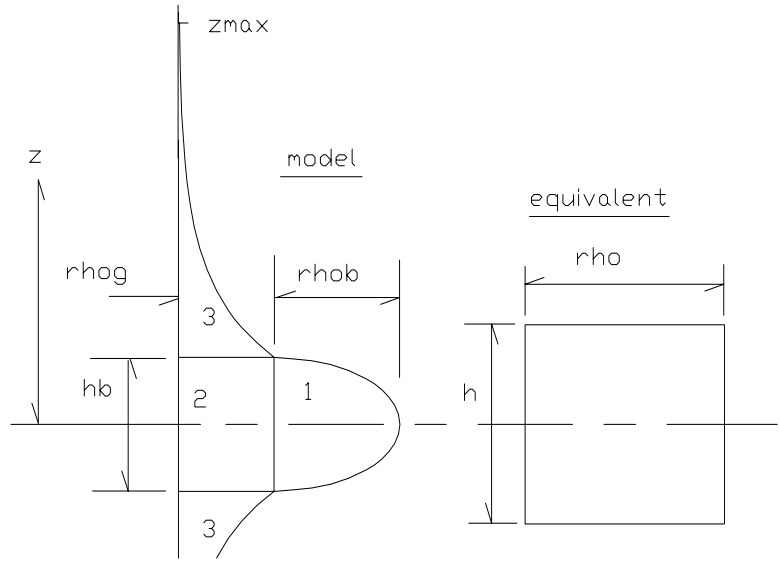

Figure 3. Model and equivalent fundamental segments

The model can be adjusted to fit conditions found with a galaxy by setting the parameters zmax / hb and f, where f is the fraction of gas / dust in the total segment mass (or SMD). The following relations result:

$$a = \frac{2z}{hb} - 1, \quad k = \frac{-\ln(.01)}{a\max}, \quad rhog = \frac{\pi}{4}\left(\frac{k}{k+0.99}\right)\left(\frac{f}{1-f}\right)rhob \qquad (2)$$

$$rho1 = rhob\sqrt{1-\left(\frac{2z}{hb}\right)^2}, \quad rho2 = rhog, \quad rho3 = rhog\ e^{-ka}, \quad rho3 = 0 \ \text{for abs}(z) > z\max$$

Unfortunately I have found no easy way to define the equivalent segment once the model segment has been set. The way I did it was to adjust h for the equivalent segment by repeated trials until the acceleration patterns for a test mass near the rim of a single ring matched those of the model. This requires integration of the effects of all parts of the model segments around the ring. However the model and equivalent I chose should be adequate for most spiral galaxies. These are:

$\quad$ zmax / hb=3, $\ f = 2/3$, $\ $ amax = 5, $\ $ k = 0.921034, $\ $ h = 1.491 hb

$\quad$ rhob=0.6329 rho, $\ $ rhog = 0.4791 rho

And the fractions of the total mass for each part become:

$\quad$ fr1, fr2, fr3 = 0.3333, 0.3213, 0.3454

The comparison curves of figure 4 are normalized by dividing by the Kepler acceleration at the ring. Both curves have a value of about one at the radius of the segments.

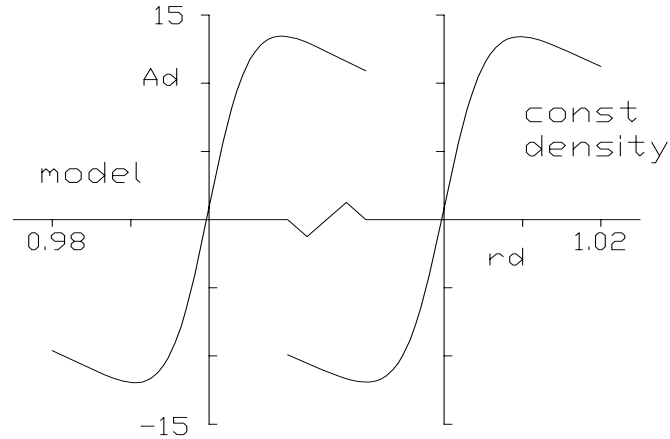

Figure 4. Test-mass accelerations for model and equivalent segments with a 5% thick single ring

Since the acceleration caused by the equivalent constant-density segment is analytic, the forward problem is simply a matter of adding up the effects of all the segments.

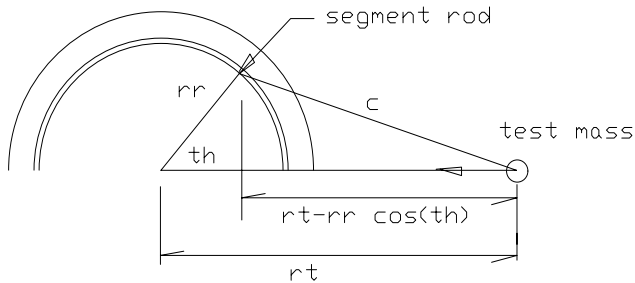 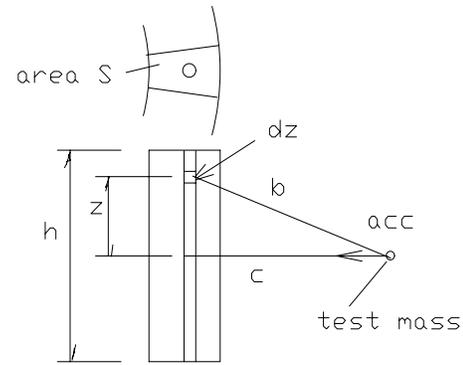

Figure 5. Geometry        Figure 6. Constant-density segment

For a single segment, acceleration of the test mass is

$$acc = \frac{G\ ddm}{c\ \sqrt{c^2 + (h/2)^2}}$$

The digital form of the equation for orbital speed at the test mass radius rt becomes:

$$v^2 = (pytks\ )^2 \times rt \times \sum_{1}^{Nr} 2 \sum_{1}^{180} \frac{G\ ddm}{\sqrt{c^2 + (h/2)^2}} \times \frac{(rt - rr\ \cos(th))}{c^2} \quad (3)$$

where $c^2 = (rr\ \sin(th))^2 + (rt - rr\ \cos(th))^2$, $pc^2$, note that c is never zero

    ddm = mass of the fundamental segment, rho r dth dr h, msuns
    dth = 1 degree, note that th = (i-1/2) dth, for i = 1 to 180
    G = gravitational constant, 4.498E-15 $pc^3$ / (msuns / $yr^2$)
    h = galaxy thickness at radius r, pcs
    Nr = number of rings

```
pytks = 9.778E5  (kms/sec)/(pcs/yr)
r    = radius to centerline of ring, pcs
rho  = density, msuns/pc³
rm   = radius to outer edge of ring, pcs
rr   = radius to rod used to represent fundamental segment mass, pcs
     = rm -dr / 2 × (rm - 0.575  dr) / (rm - dr / 2)
rt   = radius to test mass, pcs
v    = orbital speed of test mass, kms/sec
```

The reverse problem is done by repeated applications of the forward problem  Using the equivalent thickness distribution, an arbitrary density distribution is input, and the speed profile computed using (3). The speed errors (measured minus computed) at each of the ring outer radii are then used to correct the densities of the rings and the process repeated until all speed errors are small enough to be neglected.

$$\text{errv} = (vm-v) / vmax, \quad f = 0.75 \ \text{errv}, \quad \text{if all errv} < 1E-6 \text{ then quit} \quad (4)$$

$$\text{limit abs}(f) < 0.5, \quad \rho(i) = (1+f) \ \rho(i-1) \text{ for each cycle } i$$

where vm = measured speed, and vmax = maximum computed speed

After computing is done in dimensioned form, then all output data are made dimensionless by dividing with normalizing parameters:

```
accd  = acc / acckep , where acckep = GM / rmax² , the Kepler acceleration at the rim
md    = m(r) / M, where m = mass inside r, and M = total galaxy mass
hd, rd, rrd, rtd  =  (h, r, rr, rt ) / rmax, where rmax = galaxy rim radius
rhod  = rho / rhoav,  where rhoav = M / (total volume using equivalent segments)
SMDd  = SMD / SMDav,  where SMDav = M / (π rmax² )
rSMDd = rd × SMDd
vd, vmd = ( v, vm ) / vkep , where vkep = sqr(GM / rmax) , the Kepler speed at the rim
```

## 3. Checks with theory

The first results of any new method have to be proof of the equations and coding by comparison with theory. A sphere and a single ring are used for this. Equation (1) will also be used as a cross check in section 4. The solution for the rotation speeds of a test mass moving out from the center of a constant density sphere is $v = \text{sqr}(G \rho\ 4/3\ \pi)\ r$. To normalize this it is divided by the Kepler value at the rim, which is $vkep = \text{sqr}(G \rho\ 4/3\ \pi)\ rmax$. So the dimensionless form for the sphere becomes $vd = rd$. Very good results can be obtained with 100 rings, and fair results using as low as 5 rings. With some galaxies there is a spherical central bulge that might need to use as few as 5 rings to reduce the total number of rings used.  Comparisons are:

| rd  | vd (5 rings) | vd (100 rings) |
|-----|--------------|----------------|
| 0.2 | 0.1963       | 0.2000         |
| 0.4 | 0.3923       | 0.3999         |
| 0.6 | 0.5880       | 0.5999         |
| 0.8 | 0.7848       | 0.7998         |
| 1.0 | 0.9907       | 0.9978         |

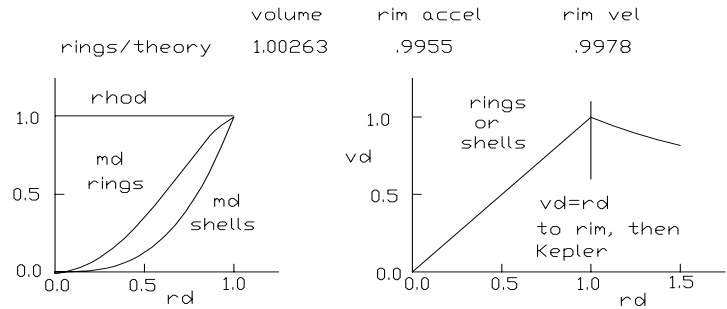

Figure 7. Comparison with theory, 100 ring sphere

Integrating from the center out in the sphere, the mass buildup is much faster for the method using rings as compared to spherical shells (figure 7). However for the rings the acceleration toward the center, caused by rings inside r, is reduced by rings outside r, to make the speed results at r for rings and shells the same.

Equation 2-146 of B & T can be used with $z = 0$ to find the results for a single ring made up of a thin zero-thickness "wire." Using the notation of this paper:

$$k^2 = \frac{4 \, r \times rt}{(r+rt)^2}, \quad \frac{1}{4}\left(\frac{k^2}{1-k^2}\right)\left(\frac{r}{rt} - \frac{rt}{r}\right) = \frac{r+rt}{r-rt}$$

and $M = \Sigma(r) \, 2\pi \, r \, dr$, then

$$acc = \frac{v^2}{rt} = \frac{G}{rt}\left[\frac{K(k)}{(r+rt)} - \frac{E(k)}{(r-rt)}\right]\frac{M}{\pi} \; : \quad \text{to normalize, acckep} = G\frac{M}{r^2}$$

and the dimensionless theoretical form for a thin-wire single ring becomes

$$accd = \frac{1}{\pi \, rtd}\left[\frac{K(k)}{(1+rtd)} - \frac{E(k)}{(1-rtd)}\right], \quad \text{with } k^2 = \frac{4 \, rtd}{(1+rtd)^2} \tag{5}$$

To coarse scale with the 5% ring the method of this paper stays close to thin-wire theory, but diverges to stay finite in fine scale, as it should.

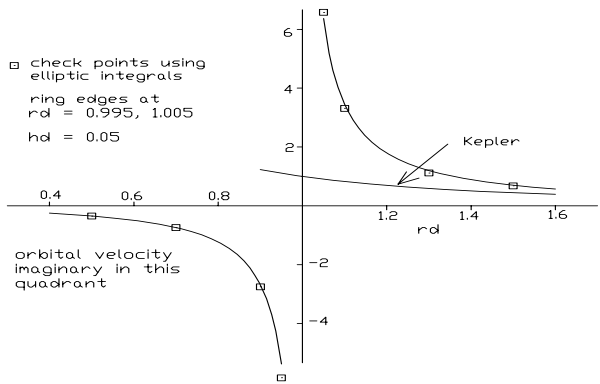
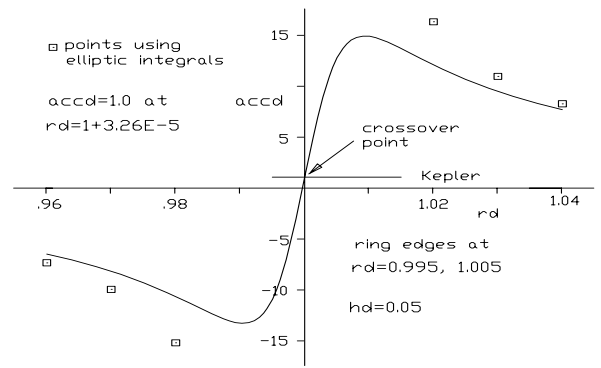

Figure 8a. Single ring, coarse scale        Figure 8b. Single ring, fine scale

As the ring thickness becomes less, the results come closer to the thin-wire theory, as expected.

|  | hd = 0.05 | hd = 0.03 | hd = 0.01 |
|---|---|---|---|
| max accd | 14.93 | 20.26 | 29.04 |
| min accd | −13.42 | −18.66 | −27.45 |

The fine-scale figure for the single ring shows the normalized acceleration is exactly one slightly outside the mid radius of the ring, just inside the circle of segment rods at rrd. The rods are at rtd = 1 + 3.8E−5, and the crossover is at rtd = 1 + 3.26E−5. Also a simple check for the location of the crossover can be done by considering a fine-wire ring with a gap at the location of a test mass. This case reduces to:

$$acc = 2G \int_\alpha^\pi \frac{dM}{c^2} \frac{c/2}{r} = 2G \frac{M}{2\pi} \int_\alpha^\pi \frac{2\,d(th/2)}{(2r)^2 \sin(th/2)} = \frac{GM}{2\pi r^2} \int_{\alpha/2}^{\pi/2} \frac{du}{\sin u}$$

$$= \frac{GM}{2\pi r^2}\left(-\ln|\tan(\alpha/4)|\right)$$

for $acc = GM/r^2$, $\ln|\tan(\alpha/4)| = -2\pi$, $\alpha = 4\tan^{-1} e^{-2\pi} = 0.00747$ rad, (6)

and the gap = $2\alpha$

So the gap is small enough to show that at the ring, the acceleration is close to the Kepler value. These ring results show that for a cloudy ring the acceleration just outside the mid radius of the ring should be about the same as if all its mass was at the center.

This completes the checkout. Equations and coding are declared correct to reasonable accuracy.

**4. Forward problem**

To get familiar with trends of the results caused by changing the mass distribution, a series of "galaxy" shapes was done, as in figure 9.

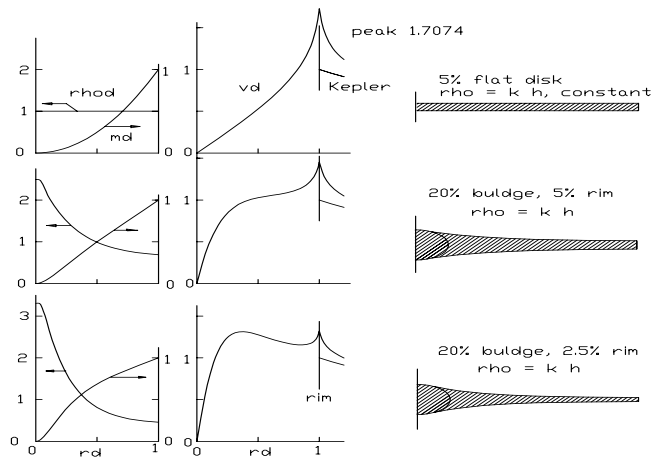

Figure 9. Trial shapes

The flat disk was a surprise. Several authors have assumed that the rotation curve for galaxies could be generated as if the acceleration at r was based only on the mass inside r acting at the center, and the mass outside had no effect, ie like a sphere. Then we would have acc(r) = G rho $\pi$ h, constant! Dimensionless acceleration would be one for all radii. This is clearly impossible because of the sudden discontinuity at the center of the disk. The dimensionless speed would be vd = sqr (rd) with a maximum of vd = 1 at the rim.

The correct curve for the 5% flat disk in figure 9 shows the speed increasing sharply toward the rim to a peak of vd = 1.7074, much larger that expected. All three "galaxies" show this spike in the speed curve, and it is caused by the thick rims. As the mass is shifted toward the center the speed curves become more Keplerian, as expected.

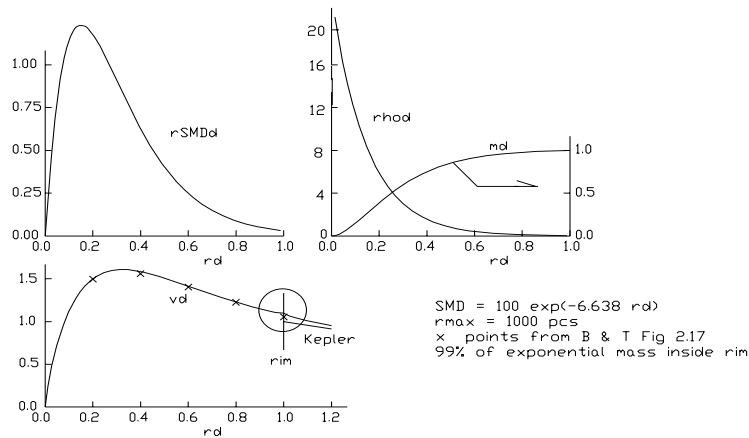

Figure 10. Exponential galaxy to match equation (1) and figure 2.17 of B & T

Equation (1) that has been used so often by the community can be checked by this method, now that other theoretical checks have been made. And of course it will also provide another check for this method. Figure 10 used hd = 0.01 to have a very thin disk almost matching that used in the reference. It shows a good match for vd vs rd except for the point at rd = 1.0. These points were read from the figure in B & T and not computed. The point at rd = 1.0 is low compared to this method because the exponential was cut off to keep 99% of the total exponential mass inside rd = 1.0, which sets rmax at $R / R_d$ = 6.638 for B & T figure 2.17. Mass was computed from B & T equation (2-170). That creates a blunt rim and a slight rise in speed. If the rest of the mass that was cut off outside rd = 1 was added, the outward attraction would reduce the speed slightly near there.

## 2. Reverse problem

Galaxies with rotational speeds constant to large radii are common, but so far past solutions for these are doubtful at best using dark-matter halos. So the first example is a constant-speed "galaxy", to show it can be done easily. For this a "Milky Way" thickness distribution is used. It was obtained from a picture in a Bok article (1981), and seems to give reasonable results for densities. This example required about 220 iteration cycles to converge. The speeds resulting from the mass distributions for all these examples are essentially identical to the input speeds. Results are in figure 11.

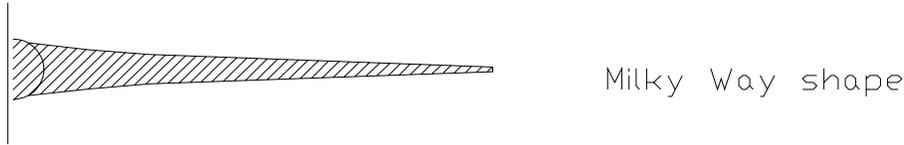

$$hd = 0.1288 \, \text{sqr}(1 - (rd/0.0644)^2) \text{ , for } rd < 0.0322 \text{ ,}$$
$$\text{then} \quad hd = 0.11158 \exp(-2.315 \, (rd - 0.0322)) \text{ , for } rd < 0.2857 \text{ ,}$$
$$\text{then} \quad hd = 0.0620 - 0.0727 \, (rd - 0.2857) \tag{7}$$
$$\text{where} \quad hd = h / rmax$$

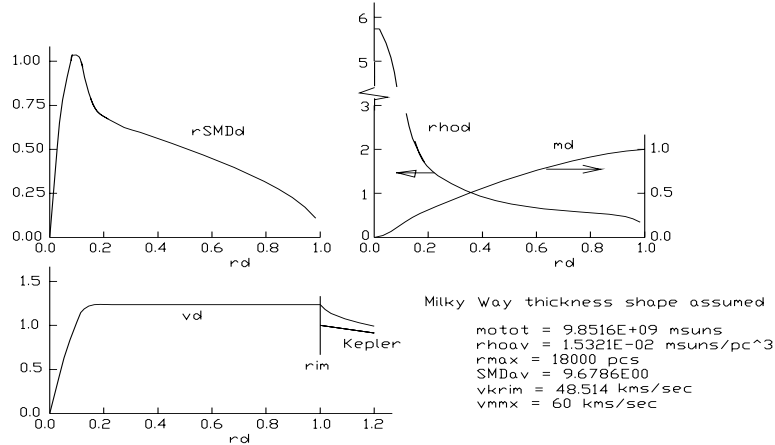

Figure 11. "Galaxy" with constant rotational speeds

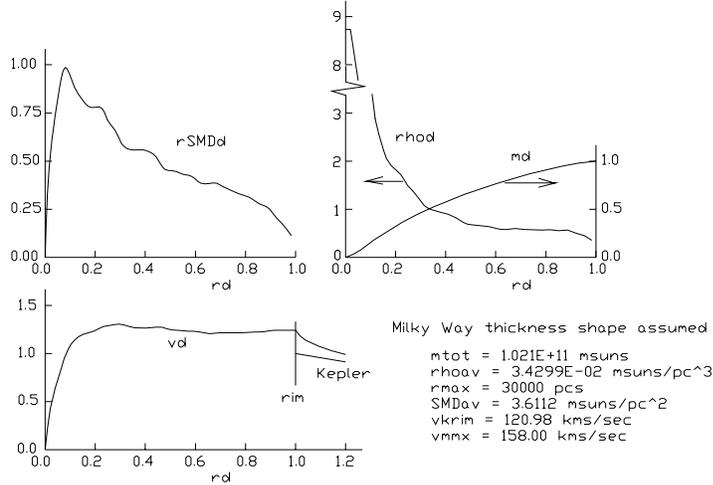

Figure 12. NGC 3198, a real galaxy with near-constant rotational speeds

NGC 3198, has almost constant rotational speeds from 6 to 30 kpcs and the results (figure 12) are similar to those of figure 11, except for the bumpiness caused by the variations in measured speeds and a reduced "spike" in rSMDd near the center. The speeds used are the same as those of figure 1, using the average values of the measurements and fairing to the test mass locations. The size of the spike is related to the smoothness of the transition of the speed curve from ramp to constant speed. The large differences in rmax and mtot don't affect the shape of the dimensionless plots. For the constant-speed range, no shear occurs between rings. The total mass shown for NGC 3198 is the most accurate so far obtained because it is based on a method that has been thoroughly checked, and has a well proven physical basis.

This next "galaxy" does not seem to occur naturally, but there is no dynamic reason for it not to exist. It is a constant-thickness disk with constant angular velocity. That is, the speeds are proportional to radius, and again no shear can occur between rings. It is equivalent to a solid flywheel, but only gravity supplies the tension to hold it together. There appears to be no known analytic solution for the mass distribution, but using this method, the SMD and density are easily obtained (figure 13). The curves show the SMD and density are constant near the center, falling off to small values near the rim. A result using constant density and solving for thickness would have a slightly different curve for rSMDd.

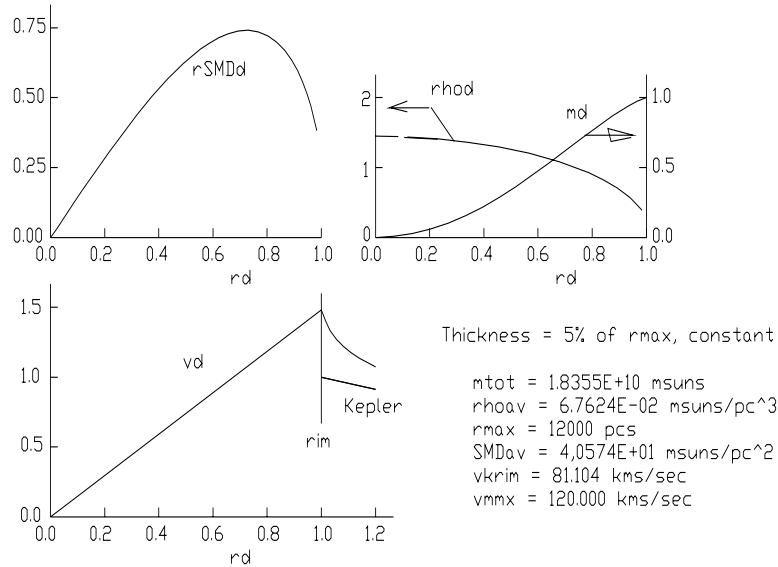

Figure 13. Constant angular velocity, constant 5% thickness, find density vs r

The final example is the Milky Way itself, using the Milky Way thickness distribution (7). The speeds are from the Bok article (1981) and they may have been changed by now, but they serve for this purpose. Some new answers can be drawn from the results.

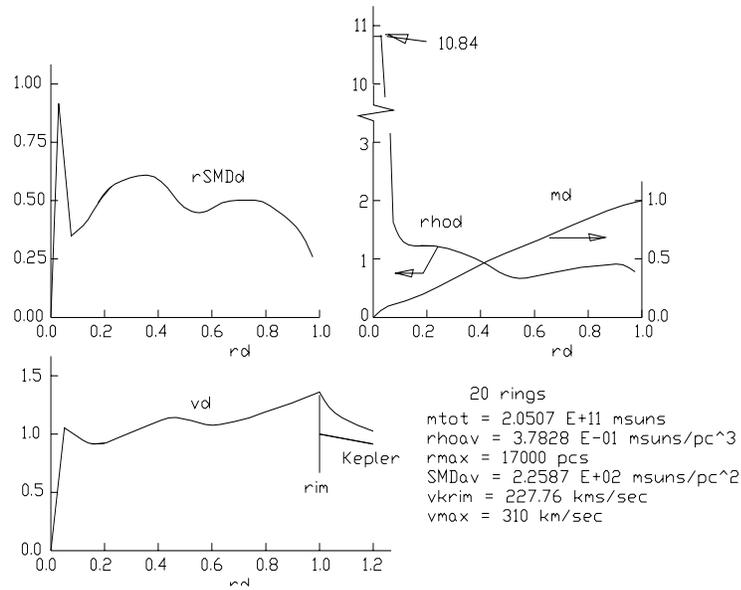

Figure 14. Milky Way

Again there is a reliable answer for the total mass of a galaxy, as good as the dimensions and measured speeds. The value of 2.0507 E11 msuns is at least a factor of 3 less than those found using equation (1) and dark-matter halos. No measured speeds are available inside rd = 0.05, so a straight line is assumed from zero to rd = 0.05. The line does not affect the computations since with only 20 rings, the first speed used is at rd = 0.05.

Inside rd = 0.05, the average density is found to be $10.84 \times 0.37828 = 4.1005$ msuns / $pc^3$, still showing an average distance between sun-sized stars of $(1 / 4.1005)^{(1/3)} = 0.625$ pcs or 2.04 light years, and much more if larger stars and some gas / dust is assumed. This is plenty of clearance between stars, so perhaps the center of our galaxy is a large elliptic collection of stars and not a black hole.

At the sun location (rd = 8.5 / 17 = 0.5), the density shown here is about 0.7 rhoav = 0.26 msuns / $pc^3$, this is much higher than 0.18 shown in B & T page 16, but at least in the same ballpark. Using a fraction of stars to total of 0.044 / 0.18 = 0.244 as in B & T page 16, the density of 0.26 leads to 2.507 pcs between sun-sized stars, or 8.2 light years on average. Since our nearest star is 4 light years away this seems about right.

## 5. Conclusions

Newton's law needs no correction and no dark-matter halos are needed to find galaxy mass distributions from rotation profiles.

Given reasonable estimates of galaxy dimensions, including thickness, good values for the mass, SMD, and density distributions are easily found from the rotation profiles.

The total masses of galaxies found using dark-matter halos are far too high.

Based on reasonable values for dimensions and their rotation profiles, the best values for the total masses of the Milky Way and NGC3198 are 2.051E11 and 1.021E11 msuns respectively.

## Acknowlegements

Binney and Tremain, with their book Galactic Dynamics, have been a great help, even though I disagree with some of the content. It's the best book I've ever read on the subject and a great source of basic data.

NASA and their Hubble telescope put the whole subject on display. The pictures almost write the equations.